\documentclass{amsart}
\usepackage[hidelinks]{hyperref}
\usepackage{algorithm}
\usepackage{algorithmic}
\usepackage{geometry}
\newtheorem{theorem}{Theorem}
\newtheorem{proposition}{Proposition}
\newcommand{\iid}{\overset{\text{i.i.d.}}{\sim}}  
\newcommand{\tHZero}{\ensuremath{\mathrm{H_0\colon F = G}}}
\newcommand{\tHOne}{\ensuremath{\mathrm{H_1\colon F \neq G}}}
\newcommand{\cHZero}{\ensuremath{\mathrm{H_0\colon AUC = 0.5}}}
\newcommand{\cHOne}{\ensuremath{\mathrm{H_1\colon AUC \neq 0.5}}}
\newcommand{\cHZeroAZero}{\ensuremath{\mathrm{H_0\colon AUC = A_0}}}

\begin{document}
\title{Wilcoxon-Mann-Whitney Test of No Group Discrimination}
\author{M. Grend\'ar}
\address{1~Laboratory of Bioinformatics and Biostatistics, Biomedical Centre Martin,
Jessenius Faculty of Medicine, Comenius University in Bratislava, Slovakia.
2~Laboratory of Theoretical Methods, Institute of Measurement Science, Slovak Academy of Sciences, Bratislava, Slovakia.
3~Bioptic Laboratory Ltd., Pilsen, Czech Republic.}

\dedicatory{To George Judge, centenarian}

\begin{abstract}
The traditional WMW null hypothesis \(\tHZero\) is erroneously too broad. 
WMW actually tests narrower \(\cHZero\). Asymptotic distribution of the standardized 
$U$ statistic (i.e., the empirical AUC) under the correct $\mathrm{H_0}$ is derived 
along with finite sample bias corrections. 
The traditional alternative hypothesis of stochastic dominance is too narrow. 
WMW is consistent against \(\cHOne\), as established by Van Dantzig in 1951.
\end{abstract}

\maketitle

\section{Introduction}

The Wilcoxon-Mann-Whitney (WMW) test statistic \cite{wilcoxon1945individual,mann1947test} is the 
$U$-statistic $U = \sum_i \sum_j 1\{X_i < Y_j\}$. 
Its standardized version $U/(n_1 n_2)$ \cite{birnbaum1955} is the empirical AUC ($\mathrm{eAUC} = \hat{P}(X < Y)$), 
cf.~\cite{pepe2003statistical},  
connecting the familiar rank-based framework to discrimination analysis.

WMW is traditionally stated to test \(\tHZero\) against \(\tHOne\) or alternatives of stochastic dominance. 
This formulation appears in most statistical textbooks \cite{lehmann1975nonparametrics} and 
software implementations \cite{R}.  We argue this formulation is incorrect.

\section{Theoretical Evidence Against Traditional $\mathrm{H_0}$}

Consider two independent samples from $N(0, \sigma_1^2)$ and $N(0, \sigma_2^2)$ with $\sigma_1 \neq \sigma_2$. 
Here $\mathrm{F \neq G}$ due to different variances, so the traditional null hypothesis \(\tHZero\) 
is false and \(\tHOne\) holds.

However, Monte Carlo simulation demonstrates that the standardized test statistic eAUC 
concentrates asymptotically on $0.5$ - the value expected under a true null hypothesis. 
The simulation (10,000 replications, $\sigma_1 = 0.1$, $\sigma_2 = 3$, $n_1 = n_2 = m = 1000$) yields 
empirical mean $0.5000069$ and $\mathrm{SD}\,\, 0.01554$. This agrees closely with the theoretical 
asymptotic distribution $N(1/2, 1/(4m))$ derived under \(\cHZero\), which 
predicts $\mathrm{SD}\,\, = 0.01581$ for $m = 1000$.
(For the extreme variance case where $\sigma_1^2 \ll \sigma_2^2$, approaching the boundary 
$\sigma_1^2/(\sigma_1^2 + \sigma_2^2) \rightarrow 0$, the asymptotic variance simplifies to $1/4$ through the 
boundary limit of the arcsin identity: as the variance ratio approaches 
zero, $\arcsin(\sigma_1^2/(\sigma_1^2 + \sigma_2^2)) \rightarrow 0$ and $\arcsin(\sigma_2^2/(\sigma_1^2 + \sigma_2^2)) \rightarrow \pi/2$, 
yielding the total asymptotic variance of $(0 + \pi/2)/(2\pi) = 1/4$.)

The empirical AUC concentrates precisely on 0.5 despite the traditional \(\tHZero\)
being false. This creates a logical contradiction: if WMW tested distributional 
equality, the test statistic should not concentrate on its null value when 
distributions clearly differ.

Though the Gaussian example invalidates the traditional claim that WMW tests \(\tHZero\), 
it does not specify the maximal class of alternatives that WMW can detect. 
This is provided by Van Dantzig \cite{vandantzig1951} who proved that WMW is consistent against alternatives 
$P(Y < X) \neq 0.5$ for independent observations from the two populations. 
For sufficiently small significance levels, the test is consistent against no other alternatives.
This consistency characterization establishes that WMW tests \(\cHZero\), 
detecting systematic pairwise dominance rather than general distributional differences.

\section{The Most General Alternative Hypothesis of WMW}

Traditional alternative hypothesis of WMW test is \(\tHOne\), cf.~\cite{lehmann1975nonparametrics}.  
Lehmann~\cite{lehmann1975nonparametrics} notes that not all pairs of distributions $\mathrm{F \neq G}$ are of equal relevance, 
and then defines stochastic dominance as the alternative hypothesis of interest. 
This restriction is unnecessarily narrow. Van Dantzig's consistency result shows 
WMW detects any departure from $\mathrm{AUC} = 0.5$. This includes cases where distributions cross multiple times, 
violating stochastic dominance assumptions, yet systematic pairwise advantage exists. 
The correct alternative $\mathrm{H_1: AUC} \neq 0.5$ subsumes stochastic 
dominance as a special case while capturing the full range of discrimination patterns WMW can detect.

\section{Asymptotic Distribution of Empirical AUC Under the Null Hypothesis \(\cHZeroAZero\)}\label{Sect:Continuous}

Though $\frac{1}{2}$ is the most interesting value of AUC as it corresponds to no
group discrimination, in general, any value of $\mathrm{A_0 \in [0, 1]}$ can be set 
at \(\cHZeroAZero\).

\subsection{General Framework}

Consider the general setup with identically and independently distributed (i.i.d.) 
random variables taking values in $\mathbb{R}$:
$X_1, \ldots, X_{n_1} \iid F$,
$Y_1, \ldots, Y_{n_2} \iid G$.
Let $n = n_1 + n_2$, 
$\lambda_n = \frac{n_1}{n} \to \lambda \in (0,1)$. 

The target parameter is the AUC: $A = P(X < Y) = \int F \, dG$, which 
under \(\cHZeroAZero\) equals $\mathrm{A_0}$ by assumption.

The empirical AUC $\hat{A}$ is a standardized $U$-statistic with kernel $h(x,y) = {1}\{x < y\}$:
$$
\hat{A} = (n_1 n_2)^{-1} \sum_{i=1}^{n_1} \sum_{j=1}^{n_2} {1}\{X_i < Y_j\}.
$$
  
\subsection{$U$-statistic Central Limit Theorem}

From the general two-sample $U$-statistic CLT:
\begin{equation*}
\sqrt{n} (\hat{A} - A_0) \xrightarrow{d} \mathcal{N}(0, \sigma^2), \quad \sigma^2 = \frac{\zeta_1^2}{\lambda} + \frac{\zeta_2^2}{1-\lambda}, 
\end{equation*}

where the variance components come from the Hoeffding projections:
\begin{align*}
\phi_1(x) &= \operatorname{E}_Y[h(x,Y)] - A_0 = \operatorname{E}_Y[{1}\{x < Y\}] - A_0 = P(Y > x) - A_0 \\
&= (1 - G(x)) - A_0 \\
\phi_2(y) &= \operatorname{E}_X[h(X,y)] - A_0 = \operatorname{E}_X[{1}\{X < y\}] - A_0 = P(X < y) - A_0 \\
&= F(y) - A_0
\end{align*}

Thus:
\begin{equation*}
\begin{aligned}
\zeta_1^2 &= \operatorname{Var}[\phi_1(X)] = \operatorname{Var}[1 - A_0 - G(X)] = \operatorname{Var}[G(X)], \\
\zeta_2^2 &= \operatorname{Var}[\phi_2(Y)] = \operatorname{Var}[F(Y) - A_0] = \operatorname{Var}[F(Y)].
\end{aligned}
\end{equation*}

Expanding these variances:
\begin{equation}
\begin{aligned}
\zeta_1^2 &= \int G^2 \, dF - \left(\int G \, dF\right)^2, \\
\zeta_2^2 &= \int F^2 \, dG - \left(\int F \, dG\right)^2. 
\end{aligned}
\tag{1}
\label{eq:vars}
\end{equation}

Under the $\mathrm{H_0}$ assumption, $\int G \, dF = 1 - A_0$ and $\int F \, dG = A_0$, equations~\eqref{eq:vars} simplify to
\begin{align*}
\zeta_1^2 &= \int G^2 \, dF - (1 - A_0)^2, \\
\zeta_2^2 &= \int F^2 \, dG - A_0^2. 
\end{align*}

\subsection{Placement Variables}

Define the \emph{placement variables}
\begin{equation*}
U := G(X) \sim H_1, \quad V := F(Y) \sim H_2.
\end{equation*}

Since $F$ and $G$ are continuous, $U$ and $V$ are uniform on $[0,1]$ if and only if $F = G$. 
In general, they follow distributions $H_1, H_2$ on $[0,1]$ satisfying
\begin{equation*}
\operatorname{E}[U] = 1 - A_0, \quad \operatorname{E}[V] = A_0.
\end{equation*}

Then
\begin{align*}
\zeta_1^2 &= \operatorname{Var}(U) = \operatorname{E}[U^2] - (1 - A_0)^2, \\
\zeta_2^2 &= \operatorname{Var}(V) = \operatorname{E}[V^2] - A_0^2. 
\end{align*}

Hence
\begin{equation*}
\sigma^2 = \frac{\operatorname{Var}(U)}{\lambda} + \frac{\operatorname{Var}(V)}{1-\lambda}. 
\end{equation*}

\subsection{Asymptotic Distribution}

Under the restriction $\int F \, dG = A_0$:
\begin{align*}
\sqrt{n} (\hat{A} - A_0) &\xrightarrow{d} \mathcal{N}(0, \sigma^2), \\
\sigma^2 &= \frac{\zeta_1^2}{\lambda} + \frac{\zeta_2^2}{1-\lambda}. 
\end{align*}

\subsection{Consistent Variance Estimation}

Compute the empirical placement values
\begin{align*}
\hat{G}(X_i) &= n_2^{-1} \sum_{j=1}^{n_2} {1}\{Y_j \leq X_i\}, \\
\hat{F}(Y_j) &= n_1^{-1} \sum_{i=1}^{n_1} {1}\{X_i \leq Y_j\}.
\end{align*}

Since $\phi_1(x) = 1 - A_0 - G(x)$ and $\phi_2(y) = F(y) - A_0$, the empirical variance estimators are
\begin{align*}
\hat{\zeta_1^2} &= (n_1-1)^{-1} \sum_{i=1}^{n_1} (1 - A_0 - \hat{G}(X_i))^2 = (n_1-1)^{-1} \sum_{i=1}^{n_1} (\hat{G}(X_i) - (1 - A_0))^2, \\
\hat{\zeta_2^2} &= (n_2-1)^{-1} \sum_{j=1}^{n_2} (\hat{F}(Y_j) - A_0)^2,
\end{align*}
and
\begin{equation*}
\hat{\sigma}^2 = \frac{\hat{\zeta_1^2}}{\lambda_n} + \frac{\hat{\zeta_2^2}}{1-\lambda_n}
\end{equation*}
is consistent for $\sigma^2$.

Therefore,
\begin{equation*}
\hat{A} \pm z_{\alpha/2} \cdot \frac{\hat{\sigma}}{\sqrt{n}}
\end{equation*}
provides an asymptotically valid $(1-\alpha)$ confidence interval for $A_0$.

\subsection{Finite Sample Bias Correction}

Under the restriction $\int F\,dG = A_0$ but
$F \ne G$, we derive bias correction using general $U$-statistic theory.
\subsubsection{Exact Bias of the Naive Estimator}
Let $\zeta_1^2 = \operatorname{Var}(G(X))$ and
$$\hat\zeta_1^2 = n_1^{-1}\sum_{i=1}^{n_1}(\hat G(X_i)-(1-A_0))^2.$$
The denominator is $n_1$, not $n_1-1$: the sum centers each term at the fixed,
known value $1-A_0$, not at an estimated sample mean, so Bessel's correction
does not apply here.

Since $\hat G(\cdot)$ is the empirical CDF of the $Y$-sample:
$$\hat G(x) = G(x) + \Delta_2(x), \qquad \operatorname{E}[\Delta_2(x)^2] = \frac{G(x)(1-G(x))}{n_2}$$
exactly, for any $n_2$, this being the variance of a binomial proportion.
By exchangeability of $X_1,\dots,X_{n_1}$ and since $\operatorname{E}[G(X)]=1-A_0$
holds exactly under the stated restriction, the cross term vanishes and
$$\operatorname{E}[\hat\zeta_1^2] = \zeta_1^2 + \frac{1}{n_2}\operatorname{E}[G(X)(1-G(X))]$$
holds \emph{exactly}, for any $n_1, n_2$ -- no asymptotic remainder is
involved in this identity. Therefore, the finite-sample bias is exactly
$$\operatorname{Bias}_1 = \frac{\omega_1}{n_2}, \qquad \omega_1 := \operatorname{E}[G(X)(1-G(X))].$$
An analogous calculation for the second term yields, exactly,
$$\operatorname{Bias}_2 = \frac{\omega_2}{n_1}, \qquad \omega_2 := \operatorname{E}[F(Y)(1-F(Y))].$$
\subsubsection{Bias-Corrected Estimators}
Estimate $\omega_1,\omega_2$ by the plug-in principle:
$$\hat\omega_1 = \frac{1}{n_1}\sum_{i=1}^{n_1}\hat G(X_i)(1-\hat G(X_i)), \qquad
\hat\omega_2 = \frac{1}{n_2}\sum_{j=1}^{n_2}\hat F(Y_j)(1-\hat F(Y_j)).$$
Because $\hat G(X_i)(1-\hat G(X_i))$ is a nonlinear function of $\hat G(X_i)$,
$\hat\omega_1$ is itself a slightly biased estimator of $\omega_1$, with error
of a smaller order than $\omega_1/n_2$ itself; we do not further correct for
this second, smaller-order effect.

The bias-corrected estimators are
\begin{equation}
\begin{aligned}
\hat\zeta_1^{2*} &= n_1^{-1}\sum_i(\hat G(X_i)-(1-A_0))^2 - \frac{\hat\omega_1}{n_2}, \\
\hat\zeta_2^{2*} &= n_2^{-1}\sum_j(\hat F(Y_j)-A_0)^2 - \frac{\hat\omega_2}{n_1}.
\end{aligned}
\tag{2}
\label{eq:vars_bias_corr}
\end{equation}
In implementation, each corrected component is floored at a small $\epsilon>0$
to guard against overshoot at small or heteroskedastic samples.

\subsubsection{Final Variance Estimator}
Insert \eqref{eq:vars_bias_corr} into the Welch-type combination:
$$\hat\sigma^2_{\text{adj}} = \frac{\hat\zeta_1^{2*}}{\lambda_n} + \frac{\hat\zeta_2^{2*}}{1-\lambda_n}, \qquad \lambda_n = n_1/n,$$
with Welch-Satterthwaite degrees of freedom
$$\nu = \frac{(\hat\sigma^2_{\text{adj}})^2}{\dfrac{[\hat\zeta_1^{2*}/\lambda_n]^2}{n_1} + \dfrac{[\hat\zeta_2^{2*}/(1-\lambda_n)]^2}{n_2}}.$$
The degrees of freedom $\nu$ follow the standard Welch--Satterthwaite
construction, which assumes each variance component is approximately
$\chi^2$-distributed with the stated degrees of freedom; here this assumption
is not exact, and the resulting inference departs from nominal even under
homoskedasticity, and worsens with heteroskedasticity. We therefore treat
$\nu$ as a practical heuristic validated empirically.

The finite-sample confidence interval is
$$\hat A \pm t_{\alpha/2,\nu} \cdot \frac{\hat\sigma_{\text{adj}}}{\sqrt n}.$$

\subsubsection{Summary}
Whenever the true AUC is $A_0$, the empirical AUC is asymptotically normal:
$$\sqrt n(\hat A - A_0) \xrightarrow{d} \mathcal N(0,\sigma^2), \qquad \sigma^2 = \frac{\zeta_1^2}{\lambda} + \frac{\zeta_2^2}{1-\lambda},$$
with consistently estimable variance via the bias-corrected estimators above.

\section{Asymptotic Distribution of Empirical AUC Under the Null Hypothesis \(\cHZeroAZero\) in the presence of ties in data}\label{Sect:Ties}

In practice, observations on a continuous variable are recorded with finite precision, potentially creating ties.
Then, the target parameter is
\begin{equation*}
A = P(X < Y) + \frac{1}{2}P(X = Y)
\end{equation*}

Consider the two-sample empirical AUC with tie-corrected kernel:
\begin{equation*}
\hat{A} = \frac{1}{M} \sum_{i=1}^{n_1} \sum_{j=1}^{n_2} h(X_i, Y_j)
\end{equation*}
where $h(x,y) = {1}\{x < y\} + \frac{1}{2}{1}\{x = y\}$ is the mid-rank kernel and $M = n_1 n_2$.

\subsection{Hoeffding Decomposition}

The Hoeffding decomposition expresses any $U$-statistic kernel in terms of its marginal projections 
and a degenerate remainder. For the kernel $h(x,y)$ with $A = \operatorname{E}[h(X,Y)]$ the decomposition is:
\begin{equation*}
h(x,y) = A + \phi_1(x) + \phi_2(y) + \psi(x,y)
\end{equation*}
where the projection functions are:
\begin{align*}
\phi_1(x) &= \operatorname{E}_Y[h(x,Y)] - A \quad \text{(first projection onto $X$-space)} \\
\phi_2(y) &= \operatorname{E}_X[h(X,y)] - A \quad \text{(second projection onto $Y$-space)} \\
\psi(x,y) &= h(x,y) - A - \phi_1(x) - \phi_2(y) \quad \text{(degenerate remainder)}
\end{align*}

For the mid-rank kernel under \(\cHZeroAZero\):
\begin{align*}
\phi_1(x) &= P(Y > x) + \frac{1}{2}P(Y = x) - A_0 = 1 - A_0 - G(x) + \frac{1}{2}\pi^Y(x) \\
\phi_2(y) &= P(X < y) + \frac{1}{2}P(X = y) - A_0 = F(y) - \frac{1}{2}\pi^X(y) - A_0
\end{align*}
where $\pi^X(t) = P(X = t)$ and $\pi^Y(t) = P(Y = t)$ are tie probabilities.

The random variables obtained by evaluating these functions at the sample points are:
\begin{align*}
\phi_1(X) &= 1 - A_0 - G(X) + \frac{1}{2}\pi^Y(X) \\
\phi_2(Y) &= F(Y) - \frac{1}{2}\pi^X(Y) - A_0 \\
\psi(X,Y) &= h(X,Y) - A - \phi_1(X) - \phi_2(Y)
\end{align*}

The decomposition for random variables becomes:
\begin{equation*}
h(X,Y) = A + \phi_1(X) + \phi_2(Y) + \psi(X,Y)
\end{equation*}

By construction, the orthogonality conditions hold:
\begin{align*}
\operatorname{E}[\phi_1(X)] &= \operatorname{E}[\phi_2(Y)] = \operatorname{E}[\psi(X,Y)] = 0 \\
\operatorname{E}[\psi(X,Y) \mid X] &= \operatorname{E}[\psi(X,Y) \mid Y] = 0
\end{align*}

\subsection{Finite-Sample Variance Identity}

From the exact Hoeffding decomposition, the finite-sample variance identity follows:
\begin{equation*}
{\operatorname{Var}(\hat{A}) = \frac{v + (n_2-1)\zeta_1^2 + (n_1-1)\zeta_2^2}{M}}
\end{equation*}
where:
\begin{align*}
v &= \operatorname{Var}(h(X,Y)) \quad \text{(kernel variance)} \\
\zeta_1^2 &= \operatorname{Var}(\phi_1(X)) \quad \text{(first projection variance)} \\
\zeta_2^2 &= \operatorname{Var}(\phi_2(Y)) \quad \text{(second projection variance)}
\end{align*}

The goal is to find an exact finite-sample unbiased estimator of $\operatorname{Var}(\hat{A})$ 
using natural sample quantities (empirical kernel variance and row/column mean variances).

\subsection{Natural Sample Quantities}

Define the empirical kernel matrix $h_{ij} = h(X_i, Y_j)$ and the following natural sample estimators:
  
\emph{Row and Column Means}
\begin{align*}
\bar{h}_{i \cdot} &= \frac{1}{n_2} \sum_{j=1}^{n_2} h_{ij} \quad \text{(row means)} \\
\bar{h}_{\cdot j} &= \frac{1}{n_1} \sum_{i=1}^{n_1} h_{ij} \quad \text{(column means)}
\end{align*}

\emph{Sample Variance Estimators}
\begin{align*}
\hat{v} &= \frac{1}{M-1} \sum_{i,j} (h_{ij} - \hat{A})^2 \quad \text{(pooled sample variance)} \\
\hat{\zeta}_1^2 &= \frac{1}{n_1-1} \sum_{i=1}^{n_1} (\bar{h}_{i \cdot} - \hat{A})^2 \quad \text{(row-mean variance)} \\
\hat{\zeta}_2^2 &= \frac{1}{n_2-1} \sum_{j=1}^{n_2} (\bar{h}_{\cdot j} - \hat{A})^2 \quad \text{(column-mean variance)}
\end{align*}

\subsection{Exact Expectation Calculations}

The exact expectations of the natural sample quantities can be obtained using the Hoeffding decomposition.

\begin{equation*}
\operatorname{E}[\hat{\zeta}_1^2] = 
\frac{(n_2-1)\zeta_1^2 + v - \zeta_2^2}{n_2}
\end{equation*}

By symmetry (swapping roles of $X$ and $Y$):
\begin{equation*}
\operatorname{E}[\hat{\zeta}_2^2] = 
\frac{(n_1-1)\zeta_2^2 + v - \zeta_1^2}{n_1}
\end{equation*}

And, 
\begin{equation*}
\operatorname{E}[\hat{v}] = v - \frac{(n_2-1)\zeta_1^2 + (n_1-1)\zeta_2^2}{M-1}
\end{equation*}

\subsection{Linear System for Unbiased Estimator}

We seek constants $(a, b, c)$ such that:
\begin{equation*}
\operatorname{E}[a\hat{v} + b\hat{\zeta}_1^2 + c\hat{\zeta}_2^2] = \operatorname{Var}(\hat{A})
\end{equation*}
for all values of $(v, \zeta_1^2, \zeta_2^2)$;
i.e., the linear combination that is unbiased for any finite sample sizes.

\subsection{Unbiased Estimator}

Linear system is solved by $(a, b, c)$ given in the following theorem:

\begin{theorem}[Exact Finite-Sample Unbiased Variance Estimator]\label{Thm:EU}
The unbiased estimator $\widetilde{\operatorname{Var}}(\hat{A})$ of $\operatorname{Var}(\hat{A})$ is:
\begin{equation*}
{\widetilde{\operatorname{Var}}(\hat{A}) = a\hat{v} + b\hat{\zeta}_1^2 + c\hat{\zeta}_2^2}
\end{equation*}
where:
\begin{align*}
a &= \frac{-(M-1)}{M(n_1-1)(n_2-1)},\\
b &= \frac{n_2^2}{M(n_2-1)},\\
c &=\frac{n_1^2}{M(n_1-1)}
\end{align*}
and $M = n_1 n_2$.
\end{theorem}

The unbiased estimator $\widetilde{\operatorname{Var}}(\hat{A})$ of $\operatorname{Var}(\hat A)$ 
can be written as
\begin{equation*}
\widetilde{\operatorname{Var}}(\hat{A}) = \frac{n_2^2(n_1-1)\,\hat{\zeta}_1^2 + n_1^2(n_2-1)\,\hat{\zeta}_2^2 - (M-1)\,\hat{v}}{M(n_1-1)(n_2-1)}
\end{equation*}

\subsection{Asymptotic Distribution of the Empirical AUC}

The exact finite-sample variance formula provides the foundation for deriving the 
asymptotic distribution of the empirical AUC under ties. 
We use the Hoeffding decomposition to establish the central limit theorem.

The Hoeffding decomposition:
\begin{equation*}
\hat{A} - A_0 = \frac{1}{n_1 n_2} \sum_{i=1}^{n_1} \sum_{j=1}^{n_2} [\phi_1(X_i) + \phi_2(Y_j) + \psi(X_i, Y_j)]
\end{equation*}

can be rewritten as

\begin{equation*}
\hat{A} - A_0 = \frac{1}{n_1}\sum_{i=1}^{n_1} \phi_1(X_i) + \frac{1}{n_2}\sum_{j=1}^{n_2} \phi_2(Y_j) + \frac{1}{n_1 n_2} \sum_{i=1}^{n_1} \sum_{j=1}^{n_2} \psi(X_i, Y_j)
\end{equation*}

Scaling by $\sqrt n$ and using $\lambda_n = n_1/n \to \lambda$, $(1-\lambda_n) = n_2/n \to (1-\lambda)$:
\begin{align*}
\sqrt{n}(\hat{A} - A_0) &= \frac{1}{\sqrt{\lambda_n}} \cdot \frac{1}{\sqrt{n_1}}\sum_{i=1}^{n_1} \phi_1(X_i) + \frac{1}{\sqrt{1-\lambda_n}} \cdot \frac{1}{\sqrt{n_2}}\sum_{j=1}^{n_2} \phi_2(Y_j) \\
&\quad + \sqrt{n} \cdot \frac{1}{n_1 n_2} \sum_{i=1}^{n_1} \sum_{j=1}^{n_2} \psi(X_i, Y_j)
\end{align*}

\subsection{Central Limit Theorem Analysis}

\subsubsection{First Term: Sample Mean of $\phi_1$}

Since $\operatorname{E}[\phi_1(X)] = 0$ (by the orthogonality property of Hoeffding projections) 
and $\operatorname{Var}[\phi_1(X)] = \zeta_1^2$, by the standard CLT:
\begin{equation*}
\frac{1}{\sqrt{n_1}}\sum_{i=1}^{n_1} \phi_1(X_i) \xrightarrow{d} \mathcal{N}(0, \zeta_1^2)
\end{equation*}

Therefore, by Slutsky's theorem (since $1/\sqrt{\lambda_n}\to 1/\sqrt\lambda$):
\begin{equation*}
\frac{1}{\sqrt{\lambda_n}} \cdot \frac{1}{\sqrt{n_1}}\sum_{i=1}^{n_1} \phi_1(X_i) \xrightarrow{d} \mathcal{N}\left(0, \frac{\zeta_1^2}{\lambda}\right)
\end{equation*}

\subsubsection{Second Term: Sample Mean of $\phi_2$}

Similarly, since $\operatorname{E}[\phi_2(Y)] = 0$ and $\operatorname{Var}[\phi_2(Y)] = \zeta_2^2$, by the standard CLT and Slutsky's theorem:
\begin{equation*}
\frac{1}{\sqrt{1-\lambda_n}} \cdot \frac{1}{\sqrt{n_2}}\sum_{j=1}^{n_2} \phi_2(Y_j) \xrightarrow{d} \mathcal{N}\left(0, \frac{\zeta_2^2}{1-\lambda}\right)
\end{equation*}

\subsubsection{Third Term: Degenerate U-statistic}

The remainder term $\psi(X_i, Y_j)$ satisfies the orthogonality conditions. 
This makes $\frac{1}{n_1 n_2} \sum_{i,j} \psi(X_i, Y_j)$ a degenerate $U$-statistic. 
By standard $U$-statistic theory, this term contributes only $O(n^{-1})$ to the asymptotic variance and vanishes in the limit:
\begin{equation*}
\sqrt{n} \cdot \frac{1}{n_1 n_2} \sum_{i=1}^{n_1} \sum_{j=1}^{n_2} \psi(X_i, Y_j) \xrightarrow{p} 0
\end{equation*}

\subsection{Asymptotic Result}

\begin{theorem}[Asymptotic Distribution of $\hat{A}$ in Presence of Ties]
Under regularity conditions and \(\cHZeroAZero\):
\begin{equation*}
\sqrt{n}(\hat{A} - A_0) \xrightarrow{d} \mathcal{N}(0, \sigma^2)
\end{equation*}
where
\begin{equation*}
{\sigma^2 = \frac{\zeta_1^2}{\lambda} + \frac{\zeta_2^2}{1-\lambda}}
\end{equation*}
with the tie-corrected variance components:
\begin{align*}
\zeta_1^2 &= \operatorname{Var}\left[1 - A_0 - G(X) + \frac{1}{2}\pi^Y(X)\right] \\
\zeta_2^2 &= \operatorname{Var}\left[F(Y) - \frac{1}{2}\pi^X(Y) - A_0\right]
\end{align*}
\end{theorem}

\textbf{Proof:} The first two terms converge to independent normal random variables with variances $\zeta_1^2/\lambda$ and $\zeta_2^2/(1-\lambda)$ respectively, and the third term vanishes in probability. 
By independence of the $X$ and $Y$ samples and Slutsky's theorem:
\begin{equation*}
\sqrt{n}(\hat{A} - A_0) \xrightarrow{d} \mathcal{N}\left(0, \frac{\zeta_1^2}{\lambda} + \frac{\zeta_2^2}{1-\lambda}\right)
\end{equation*}

\subsection{Connection to Exact Finite-Sample Results}

The asymptotic variance $\sigma^2 = \frac{\zeta_1^2}{\lambda} + \frac{\zeta_2^2}{1-\lambda}$ is 
exactly what the finite-sample variance formula approaches as $n_1, n_2 \to \infty$.

Recall that
\begin{equation}
\operatorname{Var}(\hat{A}) = \frac{v + (n_2-1)\zeta_1^2 + (n_1-1)\zeta_2^2}{n_1 n_2}
\end{equation}

Hence, using $v=\zeta_1^2+\zeta_2^2+\operatorname{Var}(\psi)$,
\begin{equation}
n \cdot \operatorname{Var}(\hat{A}) = \frac{\zeta_1^2}{\lambda_n} + \frac{\zeta_2^2}{1-\lambda_n} + \frac{\operatorname{Var}(\psi)}{n\lambda_n(1-\lambda_n)}
\end{equation}
where $\lambda_n = n_1/n$.

Since $\operatorname{Var}(\psi)$ is bounded, the last term vanishes as $n \to \infty$. Therefore:
\begin{equation}
n \cdot \operatorname{Var}(\hat{A}) \to \frac{\zeta_1^2}{\lambda} + \frac{\zeta_2^2}{1-\lambda} = \sigma^2
\end{equation}

\subsection{Reduction to Continuous Case}

\begin{proposition}[Reduction Property]
When $\pi^X(t) = \pi^Y(t) = 0$ for all $t$ (no ties), the formulas reduce exactly to the classical continuous case:
\begin{align*}
\phi_1(x) &= 1 - A_0 - G(x) \\
\phi_2(y) &= F(y) - A_0 \\
\zeta_1^2 &= \operatorname{Var}[G(X)] \\
\zeta_2^2 &= \operatorname{Var}[F(Y)]
\end{align*}
\end{proposition}

\subsection{Practical Inference}

For large samples, approximate $(1-\alpha)$ confidence intervals are:
\begin{equation*}
\hat{A} \pm z_{\alpha/2} \cdot \frac{\hat{\sigma}}{\sqrt{n}}
\end{equation*}
where $\hat{\sigma}^2$ can be either:
\begin{enumerate}
\item The exact unbiased estimator: $\widetilde{\operatorname{Var}}(\hat{A})$
\item The asymptotic plug-in: $\hat{\zeta}_1^2/\lambda_n + \hat{\zeta}_2^2/(1-\lambda_n)$
\end{enumerate}

For finite samples, the exact unbiased estimator with appropriate degrees of freedom provides better coverage. The finite-sample confidence interval is:
\begin{equation*}
\hat{A} \pm t_{\alpha/2,\nu} \cdot \sqrt{\widetilde{\operatorname{Var}}(\hat{A})}
\end{equation*}
where the degrees of freedom are computed using the Welch-Satterthwaite formula. 

Recall the unbiased estimator is $\widetilde{\operatorname{Var}}(\hat A) = a\hat v + b\hat\zeta_1^2 + c\hat\zeta_2^2$ with
\begin{equation*}
a = \frac{-(M-1)}{M(n_1-1)(n_2-1)}, \qquad b = \frac{n_2^2}{M(n_2-1)}, \qquad c = \frac{n_1^2}{M(n_1-1)},
\end{equation*}
equivalently
\begin{equation*}
\widetilde{\operatorname{Var}}(\hat{A}) = \frac{n_2^2(n_1-1)\hat{\zeta}_1^2 + n_1^2(n_2-1)\hat{\zeta}_2^2 - (M-1)\hat{v}}{M(n_1-1)(n_2-1)}.
\end{equation*}

The Welch-Satterthwaite degrees of freedom, treating each component at its natural degrees of freedom, are
\begin{equation*}
\nu = \frac{\left(\widetilde{\operatorname{Var}}(\hat{A})\right)^2}{\dfrac{(b\hat\zeta_1^2)^2}{n_1-1} + \dfrac{(c\hat\zeta_2^2)^2}{n_2-1} + \dfrac{(a\hat v)^2}{M-1}}.
\end{equation*}

The degrees of freedom $\nu$ follow the standard Welch-Satterthwaite construction, 
which assumes each variance component is approximately $\chi^2$-distributed with the 
stated degrees of freedom; here this assumption is not exact, and the resulting inference 
departs from nominal even under homoskedasticity, tied to variability in $\nu$
itself rather than a shape mismatch. We therefore treat $\nu$ as a practical heuristic, 
validated empirically.

The finite-sample confidence interval is:
\begin{equation*}
{\hat{A} \pm t_{\alpha/2,\nu} \cdot \sqrt{\widetilde{\operatorname{Var}}(\hat{A})}}
\end{equation*}

\section{Confidence intervals for the pseudomedian via test inversion}

The pseudomedian is defined as:
\begin{equation*}
\theta = \text{median}\{X_i - Y_j : i = 1, \ldots, n_1; j = 1, \ldots, n_2\}
\end{equation*}

We seek a $(1-\alpha)$-level confidence interval for $\theta$ using the correct asymptotic theory based
on testing \(\cHZero\). The confidence interval is constructed by inverting a sequence of hypothesis tests:
\begin{equation*}
\mathrm{H_0\colon} \theta = \theta_0 \quad \text{vs} \quad \mathrm{H_1\colon} \theta \neq \theta_0
\end{equation*}

For each candidate value $\theta_0$, we test whether the data are consistent with $\theta = \theta_0$.

\subsection{Testing Strategy}
To test $\mathrm{H_0\colon} \theta = \theta_0$, we:
\begin{enumerate}
\item \textit{Shift the second sample:} Define $\widetilde{Y}_j = Y_j + \theta_0$ for $j = 1, \ldots, n_2$
\item \textit{Test discrimination:} Under $\mathrm{H_0}$, we should have $\theta = \text{median}\{X_i - \widetilde{Y}_j\} = 0$
\item \textit{Equivalent AUC test:} This is equivalent to testing $\mathrm{H_0\colon AUC}(X, \widetilde{Y}) = 0.5$
\end{enumerate}


\begin{algorithm}
\caption{Confidence Interval for Pseudomedian via Test Inversion}
\begin{algorithmic}[1]
\STATE \textbf{Input:} Samples $\{X_i\}_{i=1}^{n_1}$, $\{Y_j\}_{j=1}^{n_2}$, confidence level $1-\alpha$
\STATE \textbf{Step 1:} Compute preliminary pseudomedian estimate:
\[
  \widehat{\theta} = \text{median}\{X_i - Y_j : i = 1, \ldots, n_1; j = 1, \ldots, n_2\}
\]
\STATE \textbf{Step 2:} Determine search range using robust scale:
\begin{align*}
s &= 2 \cdot \text{MAD}(\{X_i - Y_j\}) \\
[\theta_{\min}, \theta_{\max}] &= [\widehat{\theta} - 3s, \widehat{\theta} + 3s]
\end{align*}
\STATE \textbf{Step 3:} Create grid of candidate values:
\[
\{\theta_0^{(k)}\}_{k=1}^{K} = \text{equally spaced points in } [\theta_{\min}, \theta_{\max}]
\]
\STATE \textbf{Step 4:} For each $\theta_0^{(k)}$:
\begin{enumerate}
\item[4a.] Shift second sample: $\widetilde{Y}_j^{(k)} = Y_j + \theta_0^{(k)}$
\item[4b.] Test $\mathrm{H_0\colon AUC}(X, \widetilde{Y}^{(k)}) = 0.5$ 
\item[4c.] Record $p$-value: $p^{(k)}$
\end{enumerate}
\STATE \textbf{Step 5:} Determine acceptance region:
\[
  \mathcal{A} = \{\theta_0^{(k)} : p^{(k)} \geq \alpha\}
\]
\STATE \textbf{Step 6:} Confidence interval:
\[
  \text{CI}_{1-\alpha}(\theta) = [\min(\mathcal{A}), \max(\mathcal{A})]
\]
\end{algorithmic}
\end{algorithm}

\subsection{Asymptotic Validity}
Under regularity conditions, the confidence interval based on \(\cHZero\) asymptotics has coverage probability $1-\alpha$:
\[
\lim_{n_1, n_2 \to \infty} P(\theta \in \text{CI}_{1-\alpha}(\theta)) = 1-\alpha
\]

\section{\texttt{R} implementation}
Traditional asymptotic p-values of the Wilcoxon-Mann-Whitney test are derived under \(\tHZero\) and employ
variance estimation appropriate for that hypothesis. However, as demonstrated
above, WMW actually tests \(\cHZero\).
Even in the location-shift special case where \(\tHZero\) $\Leftrightarrow$ \(\cHZero\)
the variance estimation under the $\mathrm{F=G}$ framework can differ from that
under the correct $\mathrm{AUC=0.5}$ framework, leading to different p-values.

Two variants of finite-sample inference derived under \(\cHZero\) are implemented 
in the R package \texttt{wmwAUC}: an Exact Unbiased (EU) method 
and a Bias-Corrected (BC) method. EU's derivation (cf.~Sect.~\ref{Sect:Ties}) handles 
arbitrary tie patterns via the mid-rank kernel $1\{x<y\}+\tfrac12 1\{x=y\}$
from the outset. BC's derivation (cf.~Sect.~\ref{Sect:Continuous}) was for continuous 
data without ties; its implementation in \texttt{wmwAUC} nonetheless applies the 
same mid-rank kernel, for both the point estimate and every variance component, 
extending it beyond the continuous-data setting for which it was formally derived. 
Simulation confirmed that BC's conservative behavior persists under ties 
(size remained below nominal across tie proportions from 10\% to 99.8\%, 
at $n_1 = n_2 = 15$, homoskedastic), though this extension is validated 
empirically rather than re-derived theoretically.

\subsection{Exact Unbiased (EU) method}
EU (cf.~Sect.~\ref{Sect:Ties}) estimates $\operatorname{Var}(\hat A)$ by the
finite-sample unbiased combination of Theorem~\ref{Thm:EU} and reduces
correctly to the continuous case when no ties are present. For total sample
size $n_1+n_2<20$ a studentized permutation test is used in place of the asymptotic
approximation; above that threshold, the Welch--Satterthwaite degrees of
freedom are computed from the same three variance components. 

\subsection{Bias-Corrected (BC) method}
BC (cf.~Sect.~\ref{Sect:Continuous}) corrects each placement-variance component
for its upward bias (an $O(n^{-1})$ quantity known exactly, not merely to 
leading order, cf.~Sect.~4.6.1) by subtracting a plug-in estimate
of the bias, flooring each of the two corrected components independently at a
small $\epsilon>0$ rather than reverting both to their naive values if either
alone would go negative. BC uses the resulting $\hat\sigma^2_{\text{adj}}$
as the variance estimate.

\subsection{Recommendation and key functions}
Because in Monte Carlo small-samples simulations EU retained real, effect-size-responsive power in every configuration
where BC's power collapsed to zero, while remaining approximately nominal in
size, EU is the default method in \texttt{wmwAUC\_test()}; BC remains available
as an explicit option for users who want its more conservative behavior. Key
functions:
\begin{itemize}
\item \texttt{wmwAUC\_test()}: main testing function; defaults to the EU
method, with an argument to select BC instead, and a \texttt{pseudomedian}
argument controlling whether the pseudomedian is reported 
\item \texttt{wmwAUC\_pvalue\_EU()}: wmwAUC p-values via the EU method, valid
for any tie pattern; uses a studentized permutation test when $n_1+n_2<20$ and the
finite-sample-unbiased asymptotic formula otherwise
\item \texttt{wmwAUC\_pvalue\_BC()}: wmwAUC p-values via the BC method; issues
a warning when $\min(n_1,n_2)<10$, and recommends EU in that regime
\item \texttt{wmwAUC\_pseudomedian\_ci()}: confidence intervals for the
AUC-equalizing shift (pseudomedian)
\end{itemize}

The implementation (available from~\cite{grendar2025repo}) provides statistical
inference under the correct null hypothesis~\(\cHZero\), with proper and consistent
handling of ties and finite-sample corrections.

\section{Notes}

For historical precision, Van Dantzig~\cite{vandantzig1951} formulated the alternative hypothesis 
in terminology that would today be stated as 'the Area above the Ordinal 
Dominance Graph $\neq 0.5$'. The Area above the Ordinal Dominance Graph~\cite{bamber1975area} 
corresponds to the AUC. For application of the dominance statistic in psychology, 
see~\cite{cliff1993dominance}.

Notably, Van Dantzig~\cite{vandantzig1951} paired his AUC-based alternative hypothesis 
$\mathrm{H_1\colon}$ $\mathrm{AUC}$ $\neq 0.5$
with the traditional null hypothesis \(\tHZero\). This creates a logical inconsistency, 
as \(\tHZero\) is not the proper complement to \(\cHOne\). The correct 
pairing is \(\cHZero\) versus \(\cHOne\).

The condition $P(X \le Y) \ge 1/2$ is defined in~\cite{arcones2002nonparametric} as stochastic precedence.

Under location-shift assumption, the traditional \(\tHZero\) 
and the correct $\mathrm{H_0\colon}$ $\mathrm{AUC = 0.5}$  
are mathematically equivalent. However, the asymptotic distributions 
used to derive p-values differ between the two frameworks, leading to different 
variance estimation and potentially different inference.

While this work focuses on two-sample setting, similar issues arise in k-sample extensions 
(Kruskal-Wallis test), though the correct characterization requires investigation beyond the present scope.

Recent work has begun addressing limitations of traditional nonparametric frameworks. 
Conroy~\cite{conroy2012hypotheses} highlighted misinterpretations of WMW as testing median equality, 
emphasizing its role as measuring $P(X > Y)$. In~\cite{del2025invariant}, the authors
developed measures of disagreement when stochastic dominance 
assumptions fail. The present work provides a more fundamental resolution by 
establishing the correct null hypothesis and complete characterization of WMW 
sensitivity.

%
\section*{Acknowledgments}
Valuable discussions with Claude AI Sonnet 5 (Anthropic, 2026), Kimi AI (Moonshot AI, 2026) 
and ChatGPT (OpenAI, 2026) which contributed to refining ideas presented in this work, are gratefully acknowledged.
After using these tools the author reviewed and edited the content as necessary and takes 
full responsibility for the content of the publication.

%
\bibliographystyle{amsplain} 
\bibliography{references} 

\end{document}